\documentclass[prapplied,twocolumn,superscriptaddress]{revtex4-1}

\usepackage[usenames, dvipsnames]{xcolor}
\usepackage[utf8]{inputenc}
\usepackage{hyperref}
\hypersetup{colorlinks,allcolors=blue}
\usepackage{amsmath,revsymb,amssymb}
\usepackage{graphicx}
\usepackage[colorinlistoftodos]{todonotes}
\usepackage{xcolor}
\usepackage{bbold,comment}
\usepackage{braket,ulem}
\usepackage{booktabs}
\usepackage{float}

\usepackage{color, colortbl}
\definecolor{Gray}{gray}{0.9}
\newcolumntype{g}{>{\columncolor{Gray}}l}

\DeclareMathOperator{\arccot}{arccot}

\renewcommand{\emph}{\textit}

\begin{document}
    
    \title{Experimental Certification of Sustained Entanglement and Nonlocality after Sequential Measurements}
    
        \author{Giulio Foletto}
    \affiliation{
        Dipartimento di Ingegneria dell'Informazione, Universit\`a di Padova, IT-35131 Padova, Italy}
\author{Luca Calderaro}
    \affiliation{
        Dipartimento di Ingegneria dell'Informazione, Universit\`a di Padova, IT-35131 Padova, Italy}
    \author{Armin Tavakoli}
    \affiliation{Département de Physique Appliquée, Université de Genève, CH-1211 Genève, Switzerland}
        \author{Matteo Schiavon}
    \affiliation{
        Dipartimento di Ingegneria dell'Informazione, Universit\`a di Padova, IT-35131 Padova, Italy}
    \author{Francesco Picciariello}
    \affiliation{
        Dipartimento di Ingegneria dell'Informazione, Universit\`a di Padova, IT-35131 Padova, Italy}
    \author{Ad\'an Cabello}
    \affiliation{Departamento de F\'{\i}sica Aplicada II, Universidad de
        Sevilla, E-41012 Sevilla, Spain}
    \affiliation{Instituto Carlos~I de F\'{\i}sica Te\'orica y Computacional, Universidad de
        Sevilla, E-41012 Sevilla, Spain}
    \author{Paolo Villoresi}
    \affiliation{
        Dipartimento di Ingegneria dell'Informazione, Universit\`a di Padova, IT-35131 Padova, Italy}

    \affiliation{
        Istituto di
        Fotonica e Nanotecnologie, CNR, IT-35131 Padova, Italy}
        \author{Giuseppe Vallone}
         \email{vallone@dei.unipd.it}
    \affiliation{
        Dipartimento di Ingegneria dell'Informazione, Universit\`a di Padova, IT-35131 Padova, Italy}
    \affiliation{
        Dipartimento di Fisica e Astronomia, Universit\`a di Padova, IT-35131 Padova, Italy}

    \begin{abstract}
        Entanglement is a fundamental resource for quantum information science.
        However, bipartite entanglement is destroyed when one particle is observed via projective (sharp) measurements, as it is typically the case in most experiments.
        Here we experimentally show that, if instead of sharp measurements, one performs many sequential unsharp measurements on one particle which are suitably chosen depending on the previous outcomes, then entanglement is preserved and it is possible to reveal quantum correlations through measurements on the second particle at any step of the sequence.
        Specifically, we observe that pairs of photons entangled in polarization maintain their entanglement when one particle undergoes three sequential measurements, and each of these can be used to violate a CHSH inequality.
        This proof-of-principle experiment demonstrates the possibility of repeatedly harnessing two crucial resources, entanglement and Bell nonlocality, that, in most quantum protocols, are destroyed after a single measurement. The protocol we use, which in principle works for an unbounded sequence of measurements, can be useful for randomness extraction.
    \end{abstract}
    
    \maketitle
    
    \section{INTRODUCTION}
    Entanglement is at the heart of foundational and applied aspects of quantum theory \cite{Horodecki}.
    Its paradigmatic applications include cryptography \cite{Ekert91}, teleportation \cite{Teleportation}, metrology \cite{Metro}, and device-independent quantum information \cite{Focus}.
    However, it is also a fragile resource.
    The prolonged exposure of an entangled system to spontaneous decohering influences from the surrounding environment leads to its decay and eventual disappearance \cite{Eberly, Davidovich}.
    Furthermore, entanglement can vanish due to local measurements performed on one or several of the entangled systems.
    In particular, bipartite entanglement is completely destroyed as soon as a sharp measurement (i.e., a nondegenerate projective measurement) is performed on one of the two entangled systems \cite{Fuchs}.
    For example, a sharp measurement of the spin along the $x$ direction on one of the two spin qubits in a maximally entangled state leaves the qubits in a product state. 
    Nonetheless, such entanglement-breaking measurements are commonplace in entanglement-based applications of quantum theory.
    Moreover, when applied to suitable entangled states, they typically give rise to the strongest quantum correlations in tests of Bell inequalities \cite{Bell}.
    This certifies the presence of entanglement in a device-independent manner. 
    
    Recently however, a number of works have considered the generation of entanglement-based quantum correlations in scenarios in which physical systems undergo several sequential measurements \cite{Ralph, Sasmal, Shenoy}.
    It has been found possible to perform local measurements on an entangled state such that the resulting correlations violate a Bell inequality but the postmeasurement state nevertheless remains entangled enough to make yet another Bell-inequality violation achievable \cite{Ralph}.
    Naturally, this feat is impossible with projective measurements.
    The measurements must be sharp enough to generate correlations that cannot be classically modeled, but, nevertheless, unsharp enough so that some entanglement is still preserved after the measurement to make another Bell-inequality violation possible.
    These sequential unsharp measurements have been applied in studies of incompatible observables \cite{Piacentini2016}, state tomography \cite{Calderaro2018},  contextuality \cite{Anwer2019}, and self-testing of quantum instruments \cite{Mohan2019, Miklin2019}.
    Entanglement-based protocols using them have been proposed for certifying an unbounded amount of device-independent \cite{Curchod} and one-sided device-independent random numbers \cite{Coyle}, as well as for tests of finite-memory classical systems \cite{TC18}.
    
    These advances make it relevant to develop experimental tools for sustaining entanglement over sequential measurements.
    While it has already been shown that appropriately chosen unsharp measurements do not destroy entanglement \cite{Kim2012} and that others are capable of certifying it \cite{Higgins2015, White2016}, proving experimentally that it is possible to do both things in a sequential manner remains a challenge.
    Notably, two sequential violations of the Clauser-Horne-Shimony-Holt (CHSH) Bell inequality \cite{CHSH} have been demonstrated \cite{Exp1, Exp2}.
    However, extending the sequence to three and more measurements is demanding due to the sensitivity to noise \cite{Ralph}.
    Here, we demonstrate the ability to sustain entanglement over sequential measurements in a scenario in which the measurement choices depend on the history of previously performed measurements and observed outcomes.
    Since a given sequence of measurement choices and observed outcomes determines the evolution of the original state, one is faced with the task of demonstrating sustained entanglement along every possible branch of the resulting treelike structure of possible evolutions.
    We accomplish this for three sequential measurements on an entangled state, either by observing a violation of the CHSH inequality \cite{CHSH} or with a suitable entanglement witness.
    In principle, the protocol we use works for an unbounded sequence of measurements and can be useful for randomness extraction \cite{Curchod, Coyle, Pironio2010, Acin2012, Vazirani2012, Bancal2014, Acin2016}.

    \section{THEORETICAL MODEL}
    Consider a scenario in which two separated parties, Alice and Bob, share a two-qubit maximally entangled state  $\ket{\psi_1}=1/\sqrt{2}\left(\ket{00}+\ket{11}\right)$.
    Alice performs sequential measurements on her part of the state.
    In the first step, she randomly selects one of two dichotomic observables $A_0$ and $A_1$,
    \begin{equation}
     A_{m}(\mu_1)= K^\dagger_{+1|m}(\mu_1) K_{+1|m}(\mu_1)- K^\dagger_{-1|m}(\mu_1) K_{-1|m}(\mu_1) \ ,
    \end{equation}
    where $m\in\{0,1\}$ and the Kraus operators $K_{\pm 1|m}$ are defined by:
    \begin{equation}
	    \begin{aligned}
	   	& K_{+1|m}(\mu_1)=\cos(\mu_1) \Pi_{m}^++\sin(\mu_1) \Pi_{m}^-\ ,\\
	    & K_{-1|m}(\mu_1)=\sin(\mu_1)\Pi_{m}^++\cos(\mu_1) \Pi_{m}^-\ .
	    \end{aligned}
    \label{Ameas}
    \end{equation}
    Here, $\Pi_{0}^+$ and $\Pi_{0}^-$ ($\Pi_{1}^+$ and $\Pi_{1}^-$) are the projectors onto the positive and negative eigenvectors of $\sigma_Z$ ($\sigma_X$) respectively.
    Moreover, the parameter $\mu_1\in[0,\pi/4]$ can be used to tune the sharpness of her measurement \cite{NielsenChuang}.    
    On the one end, choosing $\mu_1=0$ means that the measurement is sharp (projective) and therefore consumes all the entanglement of the shared state.
    On the other end, choosing $\mu_1=\pi/4$ means that the measurement is noninteractive ($K_{\pm 1|m}=\mathbb{1}/\sqrt2$) and therefore produces random outcomes, leaving the shared state unaltered.
    Choosing $\mu_1\in (0,\pi/4)$ corresponds to an unsharp but nevertheless interactive measurement.
    Depending on Alice's choice of measurement and her observed outcome, the postmeasurement state ends up in one of four possible configurations.
    Since it is necessarily pure, it can be written in the form 
    \begin{equation}\label{post}
    \begin{aligned}
    \ket{\psi_2} &=  \frac{K_{\pm 1|m} (\mu_1) }{\sqrt{\bra{\psi_1} K^\dag_{\pm 1|m}(\mu_1)K_{\pm 1|m} (\mu_1) \ket{\psi_1}}} \ket{\psi_1}  \\
    &= U_{A,2} \otimes U_{B,2} [\cos(\eta_2) \ket{00} + \sin(\eta_2)\ket{11}]
    \end{aligned}
    \end{equation}
    for some angle $\eta_2\in(0,\pi/4]$ that quantifies the entanglement in the state and some unitary transformations $U_{A,2}$ and $U_{B,2}$ that depend on Alice's choice of measurement and observed outcome. 
    
    In the second step in the sequence, Alice uses her knowledge of the measurement choice and the recorded outcome to apply $U_{A,2}^\dagger$ to her system.
    Then, she again randomly chooses between the measurements in Eq.~\eqref{Ameas}, with the sharpness parameter denoted by $\mu_2$.
    Again, the global state $\ket{\psi_3}$ after Alice's second measurement can end up in one of four possible configurations (given knowledge of the postmeasurement state after the first step of the protocol) and it can again be written on the form of Eq.~\eqref{post}, with suitable angles and unitary operations.
    
    Acting in analogy with the second step, Alice can indefinitely continue the protocol and hence perform an arbitrarily long sequence of measurements.
    At the generic step $k$, the state is described by
    \begin{equation}
    \label{eq:starting}
    \ket{\psi_{k}} =  U_{A,k}\otimes  U_{B,k} [\cos(\eta_{k}) \ket{00} + \sin(\eta_{k})\ket{11}]\ .
    \end{equation} 
    In Table \ref{table:alphabeta} (Appendix \ref{sec:DetailedProtocol}), we give exact expressions for unitary operations $U_{A,k}$, $U_{B,k}$ and parameter $\eta_k$, which depend on the history of Alice's measurements and outcomes up to step $k-1$. 
    Alice applies $U^\dagger_{A,k}$ to her subsystem; she performs either measurement $A_0$ or $A_1$ with strength parameter $\mu_k$ and the state takes again the form of Eq. \eqref{eq:starting}, with $k$ replaced by $k+1$ so that step $k+1$ can begin.
We note that if Alice chooses $\mu_j >0\ \forall j \leq k$, then $\eta_{k+1}>0$, meaning that $\ket{\psi_{k+1}}$ is still entangled. Not only this: if she uses measurement $A_0$ with strength parameter $\mu_k > \arctan[\tan^2(\eta_k)]$ and finds outcome $-1$, the new entanglement parameter is $\eta_{k+1} = \arctan[\tan(\mu_k)/\tan(\eta_k)] > \eta_k$ and therefore entanglement has been amplified.

    \begin{figure}
        \centering
        \includegraphics[width=\columnwidth]{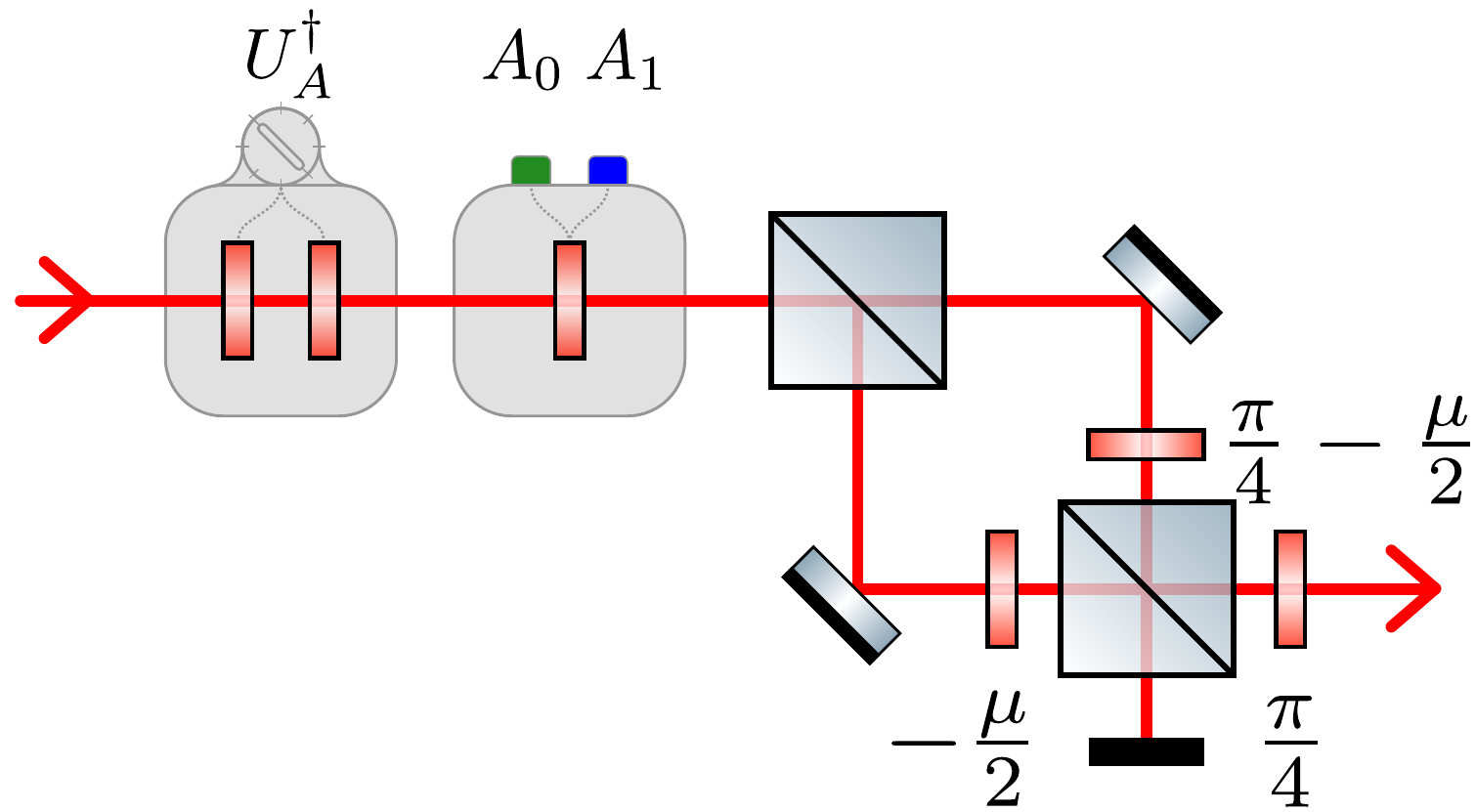}
        \caption{A conceptual optical scheme for each of Alice's steps. 
        Angle $\mu$ corresponds to the sharpness parameter in Alice's measurements. 
        The two states obtained at the two outputs correspond to the Kraus operators $K_{+1|m}$ and $K_{-1|m}$ applied to the input state, meaning that each measurement outcome is mapped to an output.
        In this model, mirrors apply the $\sigma_Z$ operation to incident polarization, whereas polarizing beam splitters simply separate two orthogonal polarizations without introducing any relative phase.
        In our implementation, one of the exits is blocked and we change the outcome corresponding to the active one by rotating the external wave plates.
        For a key to the optical elements, see Fig.~\ref{fig:Ideal}.}
        \label{fig:Block}
        \vspace*{-2mm}
    \end{figure}  
    
    \begin{figure*} [t!]
        \centering
        \includegraphics[width=0.9\textwidth]{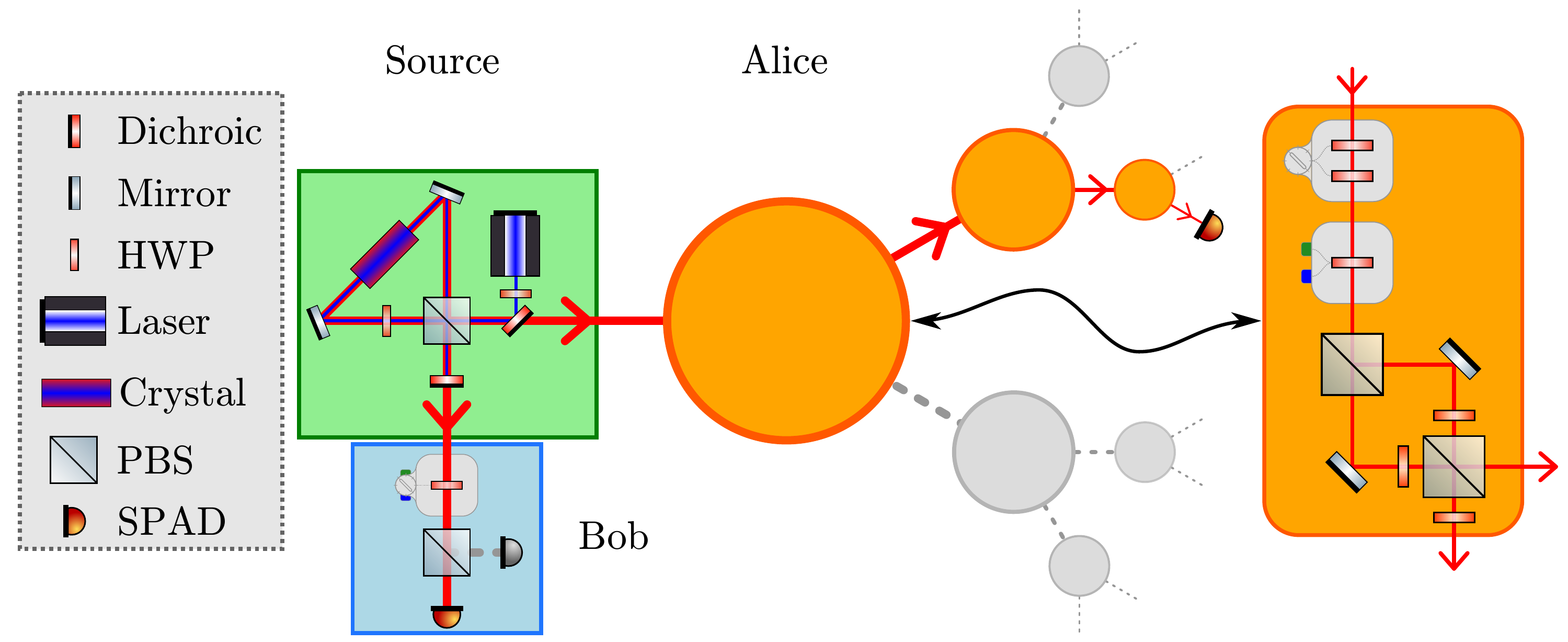}
        \caption{A conceptual treelike structure of the protocol. 
            The two positions of Bob's HWP if he stops the protocol at the kth step depend on Alice's choices and outcomes of all the previous $k-1$ steps. 
            Also, those of the HWPs that implement $U_A^\dagger$ inside each of Alice's blocks depend on her previous history.
            In our implementation, Alice stops at most at the third measurement.
        Moreover, we do not build the entire tree but only one branch and we change the combination of outcomes to which it corresponds by rotating wave plates.}
        \label{fig:Ideal}
        \vspace*{-2mm}
    \end{figure*}
    
    At any step $k$, the protocol can be interrupted for the purpose of certifying that entanglement is still present via a violation of the CHSH inequality. Bob must apply $U^\dagger_{B,k}$, projectively measure either observable $B_{0, k} = \cos(\theta_k)\sigma_X +\sin(\theta_k)\sigma_Z $ or $B_{1, k} = -\cos(\theta_k)\sigma_X +\sin(\theta_k)\sigma_Z$, where $\theta_k = \arccot[\sin(2\eta_k)]$, and finally record outcome $\pm 1$. Then, he can correlate his results with those of Alice at the same step $k$ and calculate the CHSH quantity 
    \begin{equation}\label{CHSH}
    S_{\text{CHSH}} \equiv \langle A_0B_0 \rangle + \langle A_0B_1\rangle + \langle A_1B_0 \rangle - \langle A_1B_1 \rangle.
    \end{equation}
	A violation of the CHSH inequality ($S_{\text{CHSH}}\leq 2$) certifies that the experiment involves inherently  quantum effects. With the choice of parameters that we outline, Bob finds that
	\begin{equation}
	S_{\text{CHSH,k}}= 2\cos(2\mu_k) \sqrt{1+\sin^2(2\eta_k)}\ ,
	\label{eq:CHSHResult}
	\end{equation}
	meaning that this violation ($S_{\text{CHSH,k}}> 2$) happens for any choice of $k$ if Alice uses $\mu_k<\mu_{k, max} = \frac12 \arctan[\sin (2\eta_k)]$.
    In order to find $\theta_k$ and $U_{B,k}$, Bob needs to know Alice's history of measurements and outcomes at steps $1\ldots k-1$. We can imagine that Alice feeds them back to him after each step or that he selects his operations at random from all his possibilities and then correlates his results only with those of Alice for which his choice was right. When Bob acts, he only certifies the entanglement of the state after one specific history of Alice's measurements and outcomes. However, in sufficiently many runs of the protocol, he can cover many different histories.
    
    This protocol underlines two points:
    \begin{itemize}
    	\item If they are weak, Alice's measurements do not destroy entanglement ($\mu_k >0 \rightarrow \eta_{k+1}>0$).
    	\item If they are not too weak, the same weak measurements that preserve entanglement are able to extract enough information to violate a CHSH inequality and thus certify the presence of entanglement in the premeasurement state ($\mu_k<\mu_{k, max} \rightarrow S_{\text{CHSH,k}} >2$).
    \end{itemize}
    This means that Alice can perform an arbitrary number of measurements on every single entangled system, each time fulfilling the certification requirement and without ever destroying entanglement.
    For random-number generation, she could extract more than 1 bit of certified local randomness from the sequences in which she only measures $\sigma_X$, thus beating the limit imposed by projective measurements.
    For instance, with a short sequence of two measurements, with $\mu_1 = 0.13$ and  $\mu_2 = 0$, she would certify 1.026 bits.

    \section{EXPERIMENTAL METHOD}  
    We describe here our proof-of-concept implementation aimed at verifying the two points above. Alice makes at most three sequential measurements and the protocol can be stopped at step 1, 2, or 3.
    We choose $\mu_1 \approx 0.34$, $\mu_2 \approx 0.19$ and $\mu_3 = 0$.
    The former two parameters optimize the expected values of $S_{\text{CHSH}}$ around 2.2 at steps 1 and 2, enough to grant a significant experimental observation without sacrificing the value at step 3.
    We can set $\mu_3 = 0$ because we are sure that the protocol will not continue after step 3 and therefore there is no need to preserve entanglement.
    Just before step 3, the shared system can be in 16 possible states, depending on Alice's previous choices and outcomes.
    Although a CHSH-inequality violation is possible for all of them, only in four cases is the achievable value of $S_{\text{CHSH}}$ sufficiently greater than 2, to admit the experimental detection. 
    In the remaining 12 cases, we verify entanglement using a different strategy: Alice and Bob apply the operation $U_A^\dagger \otimes U_B^\dagger$ and then measure the entanglement witness $W=\mathbb{1} \otimes \mathbb{1} - \sigma_Z \otimes \sigma_Z - \sigma_X \otimes \sigma_X$.
    It is easy to prove that the mean value of this witness is negative on the state of interest, whereas it would be positive or zero on any separable state \cite{Toth2005}.
    In total, we measure nine independent values of $S_{\text{CHSH}}$ (one when we stop the protocol at step 1, four at step 2, and four at step 3), plus 12 values of $\langle W \rangle$.
    
    We encode two qubits in the polarization degree of freedom of two separated photons.
    Polarization-entangled photon pairs are generated by a custom-built source \cite{Exp2, Calderaro2018} based on a Sagnac interferometer.
    It prepares the entangled state $\ket{\psi_1}$, where $\ket{0}$ and $\ket{1}$ refer to the horizontal and vertical polarizations.
    The pairs are sent to the two arms of our experimental setup, which correspond to Alice and Bob in the theoretical protocol.
    Fig.~\ref{fig:Block} schematizes the optical implementation for each of Alice's measurement steps: two half-wave plates (HWPs) apply $U_A^\dagger$, which is always a rotation in the space of linear polarizations; then, another HWP represents the choice between the measurements $A_0$ and $A_1$; and, finally, a polarization-based Mach-Zehnder interferometer (MZI) implements the unsharp measurement. 
    It entangles the polarization with the path degree of freedom, while the sharpness parameter is set by the angles $-\mu/2$ and $\pi/4-\mu/2$ of the internal HWPs. 
    The two exit paths correspond to the two outcomes of the polarization measurement: the probability of a photon taking each of them is equal to the probability of each outcome, while the polarization state of the photon (if observed in each path) is the expected postmeasurement state.
	One can imagine putting many of these devices in a treelike structure that, in principle, can grow unlimited, but in our experiment we stop after three of them.
	Every branch of the tree corresponds to a particular history of outcomes: detecting a photon at the end of a branch allows us to retrieve this history, attesting that the photon has taken the corresponding sequence of exits.
	Alice does not need to retrieve the outcome after each measurement, because the wave plates are set to execute the correct unitary operation $U_A^\dagger$, depending on which branch they are in.
	This encoding of measurement outcomes in the path degree of freedom is common in experiments involving sequential measurements on single photons \cite{Amselem2009, Amselem2012, Dambrosio2013, Liu2016, Zhan2017, Crespi2017}.
	Fig.~\ref{fig:Ideal} depicts this idea.
    
    Bob makes only projective measurements in the space of linear polarization; hence his scheme can be simplified to a HWP that selects the observable and a polarization beam splitter (PBS) that separates the two outcomes. 
        
    In practice, our implementation is simplified with respect to Fig.~\ref{fig:Ideal} and only uses one detector (a single-photon avalanche diode, SPAD) for Alice and one for Bob.
    Since we set Alice's third measurement to be projective, we need only two MZIs in a sequence on her side.
    One exit of each is blocked, so that there is only one path from the source to Alice's detector.
    Each interferometer is set to change the input polarization according to the Kraus operator $K_{+1|0}(\mu)=\cos(\mu) \Pi_{0}^++\sin(\mu) \Pi_{0}^-$, while two HWPs, one before and one after it, can change any of $\{\Pi_{0}^+, \Pi_{0}^-, \Pi_{1}^+, \Pi_{1}^-\}$ into another, thus selecting the basis and outcome of the measurement. This means that depending on the orientation of these plates, the interferometer can carry out each of the four Kraus operators required by the protocol.
    We mechanically rotate the HWPs in different configurations, each corresponding to one measurement-outcome combination.
    By orientating all Alice's HWPs properly, we select which branch of the tree is implemented by the one path of our setup, thus setting her complete history of measurements and outcomes. 
    The choice of measurement bases is not made in real time, photon by photon, as a faithful realization of the protocol would require, but, rather, at fixed temporal intervals, the length of which is limited by the speed of rotation of the plates and the integration time needed to keep the statistical error small enough. 
    Moreover, the setup cannot evaluate different measurement outcomes simultaneously but we have to check their relative frequencies one by one.
    We evaluate sequentially all the combinations of plate orientations, thus reconstructing the entire tree one branch at a time.
    For each combination, we count coincident detections between Alice and Bob for a fixed exposure time.
    These counts are proportional to the joint probability of obtaining the combination of outcomes under test and hence allow us to find $S_{\text{CHSH}}$ and $\langle W \rangle$.
    We note that Alice never communicates her previous history of measurements and outcomes to Bob: we externally choose it and then select the same plate orientations that Bob would use if he received such a message.
    
    We operate under the fair-sampling assumption that coincident detection events faithfully represent the photon pairs produced by the crystal.
    Moreover, our setup is affected by the ``locality loophole,'' i.e. classical communication between Alice and Bob during the measurement of $S_{\text{CHSH}}$ cannot be physically excluded.
    Finally, avoiding the tree structure increases the experiment duration, because the probabilities of outcome are not all recorded at once. Furthermore, it prevents our setup from being straightforwardly adapted for applications such as randomness extraction. However, it greatly simplifies the implementation for the goal of certifying entanglement.
    
    \section{RESULTS}
    \label{sec:results}
    We use a coincidence window of $\pm 1$ ns and an exposure time of 20 s for all measurements.
    Given the production rate of our source and the losses in the setup, the total number of photon pairs that contribute to our measurements is approximately $3\times 10^4$. 
    The detection efficiency of Alice's channel is approximately $1\%$, while Bob's is approximately $8\%$, with the difference being due to the multimode fiber on Bob's side (details in Appendix \ref{sec:DetailedSetup}).
    Before the experiment, we verify the quality of the initial entangled state using the visibility figure of merit and we obtain 99\% and 98\% when measuring the $\sigma_Z \otimes \sigma_Z$ and $\sigma_X \otimes \sigma_X$ correlations, respectively.
    The visibility in the former basis depends on the extinction ratio of the polarizing elements in the measurement setup, whereas in the latter basis it is limited by the quality of the Sagnac interferometer. 

    \begin{table}[]
        \caption{The experimental values of $S_{\text{CHSH}}$. The second column reports the history of measurements and outcomes that precede the one that yields $S_{\text{CHSH}}$ on Alice's side. The notation is as follows: outcome at step 1 $|$ measurement choice at step 1; outcome at step 2 $|$ measurement choice at step 2. The violation (i.e. $S_{\text{CHSH}}-2$) is expressed in units of the standard deviation on $S_{\text{CHSH}}$, derived from Poissonian error on the counts and error propagation.}
        \begin{tabular}{lcccc}
            \hline\hline\vspace{-0.2cm}\\
            Final & & & & Violation \\
            step & \hspace{0.2cm}Alice's history\hspace{0.2cm} & \hspace{0.2cm}$S_{\text{CHSH}}$\hspace{0.2cm} & \hspace{0.2cm}SD\hspace{0.2cm} & \hspace{0.2cm}(units of SD)\hspace{0.2cm} \\
            \hline\vspace{-0.3cm}\\
            1         & not applicable           & 2.15       & 0.01      & 20            \\
            2         & $+1|0$                   & 2.13       & 0.01      & 12           \\
            2         & $-1|0$                   & 2.07       & 0.01      & 6             \\
            2         & $+1|1$                   & 2.12       & 0.01      & 10            \\
            2         & $-1|1$                   & 2.09       & 0.01      & 7             \\
            3         & $+1|0$; $-1|0$              & 2.48       & 0.03      & 16            \\
            3         & $-1|0$; $-1|0$              & 2.53       & 0.03      & 17            \\
            3         & $+1|1$; $-1|0$              & 2.47       & 0.03      & 15            \\
            3         & $-1|1$; $-1|0$              & 2.46       & 0.03      & 15            \\
            \hline\hline
        \end{tabular}
    
    \label{tab:results}
    \end{table}
    
     \begin{table}[]
        \caption{The experimental mean values of the entanglement witness. The final step is the third for all results. The first column reports the history of measurements and outcomes that precede the one that yields $\langle W \rangle$ on Alice's side. The notation is as follows: outcome at step 1 $|$ measurement choice at step 1; outcome at step 2 $|$ measurement choice at step 2. The last column reports $-\langle W \rangle$, expressed in units of its standard deviation, derived from Poissonian error on the counts and error propagation.}
        \begin{tabular}{lccc}
            \hline\hline\vspace{-0.2cm}\\
            &&& Confirmation\\
            Alice's history\hspace{0.4cm} & 
            \hspace{0.4cm}$\langle W \rangle$\hspace{0.4cm}
            & \hspace{0.4cm}SD\hspace{0.4cm} &  \hspace{0.4cm}(units of SD)\hspace{0.4cm} \\
            \hline\vspace{-0.3cm}\\
            $+1|0$; $+1|0$              & $-0.12$       & 0.01      & 13                           \\
            $+1|0$; $+1|1$              & $-0.17$       & 0.01      & 14                           \\
            $+1|0$; $-1|1$              & $-0.20$       & 0.01      & 17                           \\
            $-1|0$; $+1|0$              & $-0.07$       & 0.01      & 8                            \\
            $-1|0$; $+1|1$              & $-0.12$       & 0.01      & 11                           \\
            $-1|0$; $-1|1$              & $-0.14$       & 0.01      & 13                           \\
            $+1|1$; $+1|0$              & $-0.06$       & 0.01      & 7                            \\
            $+1|1$; $+1|1$              & $-0.13 $      & 0.01      & 10                           \\
            $+1|1$; $-1|1$              & $-0.18 $      & 0.01      & 14                           \\
            $-1|1$; $+1|0$              & $-0.07$       & 0.01      & 8                            \\
            $-1|1$; $+1|1$              & $-0.17$       & 0.01      & 13                           \\
            $-1|1$; $-1|1$              & $-0.16$       & 0.01      & 13                           \\
            \hline\hline
        \end{tabular}
        
        \label{tab:witness}
    \end{table}

	Table \ref{tab:results} shows the experimental results for the nine values of $S_{\text{CHSH}}$, whereas Table \ref{tab:witness} shows the 12 mean values for the entanglement witnesses.
	For completeness, we also report in Table \ref{table:numvalues} (Appendix \ref{sec:numvalues}) the witnesses in the other four cases at step 3, for which the CHSH violation is a stronger certification of entanglement because it requires fewer assumptions \cite{Guhne2009}.
    We observe the violation of all the nine CHSH inequalities with several standard deviations of statistical significance, proving that Alice's sequential measurements do not destroy entanglement and at the same time can certify its presence.
    The former point is also corroborated by the results of $\langle W \rangle$, which are always significantly negative.
    We also note that the value of $S_{\text{CHSH}}$ at step 3 is greater than those at steps 1 and 2.
    This is expected given the particular sharpness parameters that we use in the experiment and proves that the protocol can be used for entanglement amplification, although only for a subset of measurement choices and outcomes.
    
    We still observe small deviations from the expected values and we attribute them to systematic alignment errors in our setup.
    Imperfections in one of Alice's interferometers might make the measurement that we perform suboptimal, thus reducing $S_{\text{CHSH}}$ at the corresponding step. 
   	Moreover, they might degrade the entanglement in the output state, thus also decreasing $S_{\text{CHSH}}$ at the steps that follow. 
   	Finally, the results at the first two steps can be influenced by defects in parts of the setup that ensue the corresponding interferometers, because photons must still go through these parts before they reach the detectors.
    The main sources of error are the phase between the arms of the MZIs, which has to be carefully regulated by tilting the PBDs, and rotation of the wave plates, which must be accurate.
    These rotations can also deviate the photons out of the detectors' entrance, thus invalidating the polarization measurements.
    Alignment difficulties such as these are the reason why simplification of the experimental setup is of paramount importance.
    Regarding statistical errors, we verify that the repeatability of the motorized rotators used to set the orientation of the wave plates is good enough that its contribution is negligible; hence the standard deviations reported in Tables \ref{tab:results} and \ref{tab:witness} are derived only from the Poissonian error on the photon counts.

    \section{CONCLUSIONS}
    In this work, we show that it is experimentally feasible to sustain entanglement over a sequence of unsharp measurements while being able to generate correlations that are strong enough to violate a Bell inequality through the same measurements.
	We report strong violations of the CHSH inequality, backed by more than 10 standard deviations of statistical significance, even at the third step of the sequence (albeit only for some of the possible histories of previous measurements and outcomes).
	This is important for protocols that require certified entanglement for quantum information tasks, such as the extraction of random bits from measurement outcomes.
    Our proof-of-principle experiment is based on entangled photon pairs and exploits only three well-controlled sequential measurements.
    It would be of evident interest to extend these ideas to other relevant physical systems that make substantially longer sequences of unsharp measurements possible, allowing one to harness entanglement many times for quantum information applications.
    
    \begin{acknowledgments}
    	Part of this work was supported by the Italian Ministry of Education, University and Research (MIUR) under the initiative ``Departments of Excellence'' (Law 232/2016). 
    	A.T. was supported by the Swiss National Science Foundation (Starting grant DIAQ, NCCR-QSIT).
    	In addition, G. F. would like to thank M. Zahidy for the useful discussions about the quantum measurement problem.
    \end{acknowledgments}
	
	\appendix
	\section{DETAILED DESCRIPTION OF THE PROTOCOL}
	\label{sec:DetailedProtocol}
	At the beginning of step $k$, Alice and Bob share the pure and entangled state
	\begin{equation}
	\ket{\psi_k} =  U_{A,k}\otimes  U_{B,k} [\cos(\eta_k) \ket{00} + \sin(\eta_k)\ket{11}]\ ,
	\end{equation}
	where $U_{A, k}$ and $U_{B, k}$ are local unitary operations and $\eta_k \in (0, \pi/4]$. 
	Alice has perfect knowledge of the state; hence she can apply $U_{A, k}^\dagger$ to her subsystem.
	The shared state becomes
	\begin{equation}
	\ket{\psi'_k} = \mathbb{1}_A\otimes  U_{B,k} [\cos(\eta_k) \ket{00} + \sin(\eta_k)\ket{11}]\ .
	\end{equation}
	She chooses the strength of her measurement, in the form of parameter $\mu_k \in (0, \mu_{k, max})$, where
	\begin{equation}
	\mu_{k, max} = \frac12 \arctan[\sin (2\eta_k)]\ .
	\end{equation}
	We require $\mu_k > 0$ to preserve entanglement at step $k+1$ (indeed, Alice is allowed to choose $\mu_k = 0$ if she agrees with Bob to stop the protocol at step $k$).
	The upper bound is required to make the violation of the CHSH inequality at step $k$ possible. 
	Note that this implies that $\tan(2\mu_k) \leq \sin(2\eta_k) \leq \tan(2\eta_k)$ and hence $\mu_k \leq \eta_k$.
	
	Then, she chooses between the two observables $A_0(\mu_k) = E_{+1|0} (\mu_k)- E_{-1|0}(\mu_k)$ and $A_1 (\mu_k) = E_{+1|1}(\mu_k)- E_{-1|1}(\mu_k)$, where
	\begin{equation}
	\begin{aligned}
	E_{+1|0}(\mu_k) &= \frac12 \left[\mathbb{1} + \cos(2\mu_k)\sigma_Z \right]\ ,\\
	E_{-1|0}(\mu_k) &= \frac12 \left[\mathbb{1} - \cos(2\mu_k)\sigma_Z \right]\ ,\\
	E_{+1|1}(\mu_k) &= \frac12 \left[\mathbb{1} + \cos(2\mu_k)\sigma_X \right]\ ,\\
	E_{-1|1}(\mu_k) &= \frac12 \left[\mathbb{1} - \cos(2\mu_k)\sigma_X \right]\ .\\
	\end{aligned}
	\end{equation}
	$A_0(\mu_k)$ and $A_1(\mu_k)$ are noisy measurements of $\sigma_Z$ and $\sigma_X$, respectively.
	Moreover, $E_{\pm 1|m}(\mu_k) = K_{\pm 1|m}(\mu_k)^\dagger K_{\pm 1|m}(\mu_k)$ where $K_{\pm 1|m}(\mu_k)$ are the Kraus operators mentioned in the main text:
	\begin{equation}
	\begin{aligned}
	K_{+1|0}(\mu_k) &= \cos(\mu_1)\ket{0}\bra{0}+\sin(\mu_k)\ket{1}\bra{1}\ ,\\
	K_{-1|0}(\mu_k) &= \sin(\mu_1)\ket{0}\bra{0}+\cos(\mu_k)\ket{1}\bra{1}\ ,\\
	K_{+1|1}(\mu_k) &= \cos(\mu_1)\ket{+}\bra{+}+\sin(\mu_k)\ket{-}\bra{-}\ ,\\
	K_{-1|1}(\mu_k) &= \sin(\mu_1)\ket{+}\bra{+}+\cos(\mu_k)\ket{-}\bra{-}\ ,
	\end{aligned}
	\label{eq:Krauses}
	\end{equation}
	where $\ket{+}$ and $\ket{-}$ are the two eigenstates of $\sigma_X$.
	
	After performing the measurement and recording the outcome, the shared state becomes
	\begin{equation}
	\ket{\psi_{k+1}} =  U_{A,k+1}\otimes  U_{B,k+1} [\cos(\eta_{k+1}) \ket{00} + \sin(\eta_{k+1})\ket{11}]
	\end{equation}
	and step $k+1$ can begin.
	
	The unitary operations $U_{A,k+1}$ and $U_{B,k+1}$ and the new parameter $\eta_{k+1}$ can be found from their corresponding values at step $k$.
	In particular,
	\begin{equation}
	\begin{aligned}
	U_{A,k+1} &= e^{-i\alpha_{k+1} \sigma_Y}\ ,\\
	U_{B,k+1} &= e^{-i\beta_{k+1} \sigma_Y} U_{B,k}\ ,
	\end{aligned}
	\end{equation}
	where angles $\alpha_{k+1}$ and $\beta_{k+1}$ depend on the choice of measurement and outcome at step $k$, as summarized in Table \ref{table:alphabeta}.
	
	\begin{table*}[]
		\renewcommand{\arraystretch}{1.5}
		\caption{The properties of step $k+1$ given those of step $k$.}
		\centering
		\begin{tabular}{lcccc}
			\hline\hline\rowcolor{white}
			Kraus  & \hspace{2.25cm}$\alpha_{k+1}$\hspace{2.25cm} & \hspace{2.25cm}$\beta_{k+1}$\hspace{2.25cm} & \hspace{2.25cm}$\eta_{k+1}$\hspace{2.25cm} \\
			at step $k$\hspace{0.4cm} & & & \\ \hline
			$K_{+1|0}$	&	0													&	0													& $\arctan[\tan(\mu_k)\cdot\tan(\eta_k)]$   \\
			$K_{-1|0}$	&	$\pi/2$												&	$\pi/2$												& $\arctan[\tan(\mu_k)/\tan(\eta_k)]$	\\
			$K_{+1|1}$	&	$\frac12 \arccot[\tan(2\mu_k)\cdot\cos(2\eta_k)]$	&	$\frac12 \arctan[\tan(2\eta_k)\cdot\cos(2\mu_k)]$	& $\frac12 \arcsin[\sin(2\mu_k)\cdot\sin(2\eta_k)]$   \\
			$K_{-1|1}$	&	$-\frac12 \arccot[\tan(2\mu_k)\cdot\cos(2\eta_k)]$	&	$-\frac12 \arctan[\tan(2\eta_k)\cdot\cos(2\mu_k)]$	& $\frac12 \arcsin[\sin(2\mu_k)\cdot\sin(2\eta_k)]$	\\
			\hline\hline
		\end{tabular}
		\label{table:alphabeta}
	\end{table*}

	We emphasize that if the measurement choice is $A_0$, the outcome is $-1$ and $\tan(\mu_k) > \tan^2(\eta_k)$, then $\eta_{k+1} > \eta_k$, which means that entanglement has been amplified; this cannot happen in the other cases.
	To find the expressions of Table \ref{table:alphabeta}, one should write $\ket{\psi_{k+1}} = K_{\pm 1|m} (\mu_k) /\sqrt{\bra{\psi'_k} E_{\pm 1|m} (\mu_k) \ket{\psi'_k}} \ket{\psi'_k}$ and then perform the Schmidt decomposition on this state.
	The singular vectors (which form the columns of $U_{A, k+1}$ and $U_{B, k+1}$) should be ordered according to decreasing singular values.
	Then, $\tan(\eta_{k+1})$ is simply the ratio between the smaller and larger singular values.
	This sequence begins at step 1 with $U_{A, 1} = U_{B, 1} = \mathbb{1}$ and $\eta_1 = \pi/4$.
	With this information and the above updating rules, it is possible to find all parameters at all steps.
	
	If Alice and Bob decide to interrupt the protocol at step $k$, Bob must apply $U_{B, k}^\dagger$ and measure projectively the two observables $B_{0, k} = \cos(\theta_k)\sigma_X +\sin(\theta_k)\sigma_Z$ and $B_{1, k} = -\cos(\theta_k)\sigma_X +\sin(\theta_k)\sigma_Z$, where $\theta_k = \arccot[\sin(2\eta_k)]$.
	Inserting these expressions in the definition of $S_{\text{CHSH}}$ yields Eq. \eqref{eq:CHSHResult}.
	From this, one can prove that in order to violate the CHSH inequality, $\mu_k$ must be chosen such that $\tan(2\mu_k)<\sin(2\eta_k)$, as stated in the main text.
	
	\section{VALUES FOR THE THREE-STEPS IMPLEMENTATION}
	\label{sec:numvalues}   
	Table \ref{table:numvalues} contains the numerical values for all the parameters of the protocol, restricted to our three-steps implementation.
	
	\begin{table*}
		\caption{The values of the parameters for our three-steps implementation and comparison with the observed values for $S_{\text{CHSH}}$ and $\langle W \rangle$. The notation for the second column is as follows: outcome at step 1 $|$ measurement choice at step 1; outcome at step 2 $|$ measurement choice at step 2. The standard deviations in the last two columns are derived from Poissonian error on the counts and error propagation.}
		\renewcommand{\arraystretch}{1.2}
		\centering
		\begin{tabular}{lcccccccccc}
			\hline\hline
			Step $k$ \hspace{0.15cm}& \hspace{0.15cm}Alice's history\hspace{0.15cm} &  \hspace{0.15cm}$\eta_{k}$\hspace{0.15cm} & \hspace{0.15cm}$\alpha_{k}$\hspace{0.15cm} & \hspace{0.15cm}$\beta_{k}$\hspace{0.15cm} & \hspace{0.15cm}$\theta_{k}$\hspace{0.15cm} & \hspace{0.15cm}$\mu_k$\hspace{0.15cm} & \hspace{0.15cm}$S_{\text{CHSH}}$\hspace{0.15cm} & 
			\hspace{0.15cm}$\langle W \rangle$\hspace{0.15cm} &
			\hspace{0.15cm}$S_{\text{CHSH}}$ (observed)\hspace{0.15cm} & 
			\hspace{0.15cm}$\langle W \rangle$ (observed)\hspace{0.15cm} \\ \hline
			1	&	not applicable	&	$\ \ \pi/4\ \ $	&	\hspace{.5cm}0\hspace{.5cm}	&	\hspace{.5cm}0\hspace{.5cm}	&	
			$\ \ \ \pi/4\ \ \ $	& $\ \ 0.34\ \ $	&	2.20	&	$-1$ & $2.15 \pm 0.01$ & no data	\\
			2	&	$+1|0$	&	0.34	&	0	&	0	&	1.01	&	0.19	&	2.19	&	$-0.63$ &	$2.13 \pm 0.01$ & no data	\\
			2	&	$-1|0$	&	0.34	&	$\pi/2$	&	$\pi/2$	&	1.01	&	0.19	&	2.19	&	$-0.63$ &	$2.07 \pm 0.01$ & no data	\\
			2	&	$+1|1$	&	0.34	&	$\pi/4$	&	$\pi/4$	&	1.01	&	0.19	&	2.19	&	$-0.63$ &	$2.12 \pm 0.01$ & no data	\\
			2	&	$-1|1$	&	0.34	&	$-\pi/4$	&	$-\pi/4$	&	1.01	&	0.19	&	2.19	&	$-0.63$ &	$2.09 \pm 0.01$ & no data	\\
			3	&	$+1|0; +1|0$	&	0.07	&	0	&	0	&	1.44	&	0	&	2.02	&	$-0.14$	&	no data	&	$-0.12 \pm 0.01$\\
			3	&	$+1|0; -1|0$	&	0.50	&	$\pi/2$	&	$\pi/2$	&	0.87	&	0	&	2.61	&	$-0.84$	&	$2.48 \pm 0.03$ & $-0.75 \pm 0.01$  \\
			3	&	$+1|0; +1|1$	&	0.12	&	0.63	&	0.32	&	1.34	&	0	&	2.05	&	$-0.23$	&	no data	&	$-0.17 \pm 0.01$\\
			3	&	$+1|0; -1|1$	&	0.12	&	$-0.63$	&	$-0.32$	&	1.34	&	0	&	2.05	&	$-0.23$ &	no data	&	$-0.20 \pm 0.01$	\\
			3	&	$-1|0$; $+1|0$	&	0.07	&	0	&	0	&	1.44	&	0	&	2.02	&	$-0.14$ &	no data	&	$-0.07 \pm 0.01$	\\
			3	&	$-1|0; -1|0$	&	0.50	&	$\pi/2$	&	$\pi/2$	&	0.87	&	0	&	2.61	&	$-0.84$ &	$2.53 \pm 0.03$ & $-0.79 \pm 0.01$	\\
			3	&	$-1|0; +1|1$	&	0.12	&	0.63	&	0.32	&	1.34	&	0	&	2.05	&	$-0.23$ &	no data	&	$-0.12 \pm 0.01$	\\
			3	&	$-1|0; -1|1$	&	0.12	&	$-0.63$	&	$-0.32$	&	1.34	&	0	&	2.05	&	$-0.23$	&	no data	&	$-0.14 \pm 0.01$ \\
			3	&	$+1|1; +1|0$	&	0.07	&	0	&	0	&	1.44	&	0	&	2.02	&	$-0.14$	&	no data	&	$-0.06 \pm 0.01$\\
			3	&	$+1|1; -1|0$	&	0.50	&	$\pi/2$	&	$\pi/2$	&	0.87	&	0	&	2.61	&	$-0.84$ &	$2.47 \pm 0.03$ & $-0.78 \pm 0.01$	\\
			3	&	$+1|1; +1|1$	&	0.12	&	0.63	&	0.32	&	1.34	&	0	&	2.05	&	$-0.23$ &	no data	&	$-0.13 \pm 0.01$	\\  		
			3	&	$+1|1; -1|1$	&	0.12	&	$-0.63$	&	$-0.32$	&	1.34	&	0	&	2.05	&	$-0.23$ &	no data	&	$-0.18 \pm 0.01$	\\
			3	&	$-1|1; +1|0$	&	0.07	&	0	&	0	&	1.44	&	0	&	2.02	&	$-0.14$ &	no data	&	$-0.07 \pm 0.01$	\\
			3	&	$-1|1; -1|0$	&	0.50	&	$\pi/2$	&	$\pi/2$	&	0.87	&	0	&	2.61	&	$-0.84$ &	$2.46 \pm 0.03$ & $-0.68 \pm 0.02$	\\
			3	&	$-1|1; +1|1$	&	0.12	&	0.63	&	0.32	&	1.34	&	0	&	2.05	&	$-0.23$ &	no data	&	$-0.17 \pm 0.01$	\\
			3	&	$-1|1; -1|1$	&	0.12	&	$-0.63$	&	$-0.32$	&	1.34	&	0	&	2.05	&	$-0.23$ &	no data	&	$-0.16 \pm 0.01$	
			\\
			\hline\hline
		\end{tabular}
	\label{table:numvalues}
	\end{table*}
	
	\section{DETAILED DESCRIPTION OF THE EXPERIMENTAL SETUP}
	\label{sec:DetailedSetup}
	The heart of our entangled-photons source is a 30-mm-long periodically poled potassium titanyl phosphate (PPKTP) crystal, which lies inside a Sagnac interferometer.
	A continuous-wave laser at 404 nm sends diagonally polarized light to the PBS of the interferometer so that the crystal is illuminated from both directions.
	By a spontaneous parametric down-conversion process in a quasiphase-matching configuration,  pairs of orthogonally polarized photons at 808 nm are generated.
	Due to a dual wavelength half-wave plate (that works both at 404 nm and 808 nm) inside the Sagnac interferometer, the quantum state just after it is $1/\sqrt{2}(\ket{01} + \ket{10})$, where the horizontal ($\ket{0}$) and vertical ($\ket{1}$) polarizations are defined by the aforementioned PBS.
	Two single-mode fibers collect the photons and bring them to the two arms of the measurement setup, Alice and Bob.
	In each of them, a HWP and a QWP correct the unitary operations applied by the fibers.
	Bob also uses a liquid-crystal retarder (LCR) to fine tune the phase between different polarization components.
	These optical elements change the state to
	\begin{equation}
	\ket{\psi_1}=\frac{1}{\sqrt{2}}(\ket{00} + \ket{11})\ ,
	\end{equation}
	where $\ket{0}$ and $\ket{1}$ are now defined by Alice's and Bob's polarizers.
	
	The principle of the two measurement setups is that polarizing and birefringent optical elements select one measurement effect and then their axes are rotated to evaluate sequentially all the effects of interest.
	The number of coincident detections in each configuration is counted and associated with the corresponding effect.
	Bob has to measure only linear polarizations; therefore his measurement setup consists of a rotating HWP and a fixed linear polarizer (LP). 
	A multimode fiber then collects the photons and brings them to a SPAD.
	On Alice's side, two Mach-Zehnder interferometers in a series implement the two weak measurements. 
	They separate the horizontal- and vertical-polarization components using PBDs.
	For convenience, we use three HWPs instead of two in our MZIs: in this way, we can regulate the sharpness parameter $\mu$ by rotating a single plate, while the others are fixed.
	Indeed, the arm carrying the $\ket{0}$ polarization encounters a HWP with an axis at $-\pi/8$, while the other encounters a HWP at $\pi/8$. 
	Then, a HWP at angle $\pi/8-\mu/2$ spans across both.
	The interferometer, followed by a HWP at angle $\pi/4$ that swaps $\ket{0}$ and $\ket{1}$ implements the Kraus operator:
	\begin{equation}
	K_{+1|0}(\mu)=\cos(\mu) \Pi_{0}^++\sin(\mu) \Pi_{0}^-\ .
	\end{equation}
	Two HWPs, one before and one after the MZI, can change any of $\{\Pi_{0}^+, \Pi_{0}^-, \Pi_{1}^+, \Pi_{1}^-\}$ into another, thus selecting the basis and outcome of the measurement.    
	This means that depending on the orientation of these plates, the interferometer can carry out each of the four Kraus operators required by the protocol, shown in Eq. \eqref{eq:Krauses}.
	
\begin{figure}
		\centering
		\includegraphics[width=0.95\linewidth]{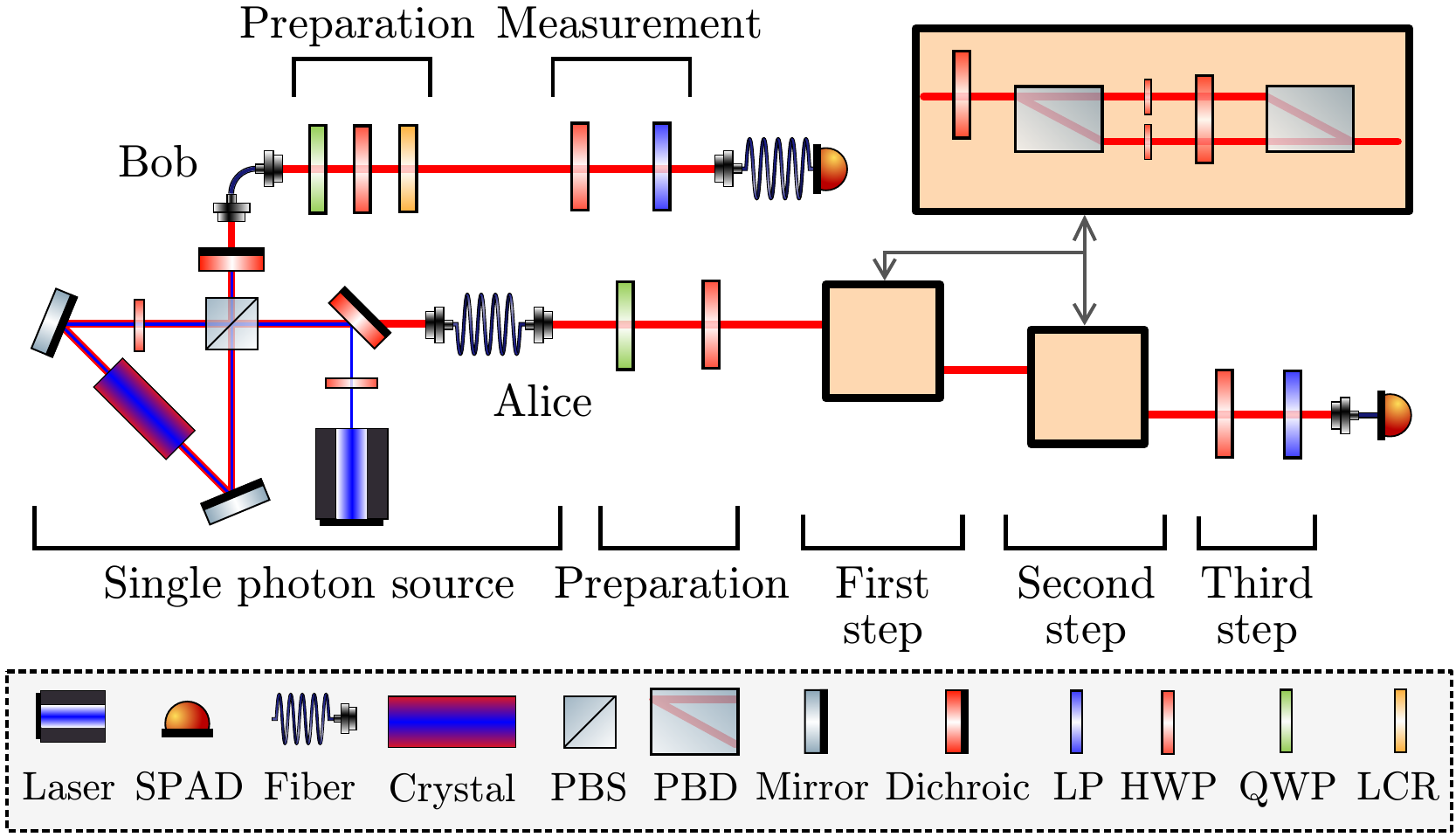}
		\caption{The actual optical implementation.}
		\label{fig:Actual}
		\vspace*{-2mm}
	\end{figure}

	The unitary operations needed before each weak measurement are realized by the HWP at the beginning of the next step.
	The total number of HWPs needed between the measurement steps would be five (one to select the measurement-outcome combination of the previous measurement, one to do the same for the next one, one to swap $\ket{0}$ and $\ket{1}$, and two for the unitary operation), but this can be reduced to one, as is true for any odd number of HWPs.
	Since the third measurement is strong, it is achieved by a HWP and a LP.
	A single-mode fiber finally collects Alice's photons and brings them to a SPAD, the signal of which is correlated with Bob's signal by a time-tagger with 80-ps resolution, that then returns coincidence counts within a $\pm 1$-ns window.
	A faithful representation of our implementation is shown in Fig.~\ref{fig:Actual}.
	
	The total rate of coincidences summed over the outcomes of a polarization measurement is about 1500 Hz.
	We measure the efficiency of Alice's detection system as the ratio between the rate of coincidences and that of single counts in Bob's channel and we obtain approximately $1\%$. 
	This value includes the quantum efficiency of Alice's SPAD and losses in the optical system, but is mostly limited by the two couplings into single-mode fiber that photons must endure in their path from the crystal to the detector. 
	Bob's efficiency is comparatively much better (approximately $8\%$) because of the multimode fiber (with a higher collection probability) that we use at the detection stage.

	The coincidence rate, integrated over an exposure time of 20 seconds, makes the total number of coincident events contributing to a complete measurement about $3\times 10^4$, sufficient to make statistical errors small.
	Systematic misalignments of the setup are the main source of error.
	In particular, imperfections in the wave plates can cause imbalances in the photon counts, which are critical for the final results.
	Rotating plates can slightly deviate the beam out of the fiber entrance, hindering the accuracy of the polarization measurements.
	Preparation of the entangled state is also important and needs precise alignment of the source.
	Finally, the Mach-Zehnder interferometers need to be perfectly balanced to achieve sufficient visibility.

    \end{document}